\documentclass[aps,prl,twocolumn]{revtex4-2}%
\usepackage{hyperref}
\usepackage[]{graphicx}

\usepackage{xcolor}%
\newcommand{\chem}[1]{\ensuremath{\mathrm{#1}}}

\begin{document}
	\preprint{APS/123-QED}
	
	\title{Adiabatic demagnetization refrigeration to mK temperatures with the distorted square lattice magnet NaYbGeO$_{4}$}
	
	\author{U. Arjun$^{1,2}$}
	\email{arjunu@iisc.ac.in}
	
	\author{K. M. Ranjith$^3$}
	
	\author{A. Jesche$^2$}

	\author{F. Hirschberger$^2$}
	\author{D. D. Sarma$^1$}
	\author{P. Gegenwart$^2$}
	\email{philipp.gegenwart@physik.uni-augsburg.de}
	
	\affiliation{$^1$Solid State and Structural Chemistry Unit, Indian Institute of Science, Bangalore-560012, India}
	\affiliation{$^2$Experimental Physics VI, Center for Electronic Correlations and Magnetism, Institute of Physics, University of Augsburg, 86135 Augsburg, Germany}
	\affiliation{$^3$Laboratoire National des Champs Magn\'{e}tiques Intenses-EMFL, CNRS, Université Grenoble Alpes, 38042 Grenoble, France}

	\date{\today}
\begin{abstract}
We report the synthesis, characterization, low-temperature magnetic, and thermodynamic measurements of the novel milli-Kelvin adiabatic demagnetization refrigeration (mK-ADR) candidate material NaYbGeO$_4$ which exhibits a distorted square lattice arrangement of YbO$_{6}$ magnetic units. Magnetization and specific heat indicate weakly interacting effective spin-1/2 moments below 10~K, with a Curie-Weiss temperature of only 15~mK, that can be polarized by magnetic fields of order 1~T. For the ADR performance test, we start the demagnetization from 5~T at a temperature of $\sim 2$~K and reach a minimum temperature of 150~mK at zero field. The warming curve indicates a sharp magnetic transition in the heat capacity at 210~mK, implying only weak magnetic frustration. The entropy density of $S_{\rm GS}\simeq 101$ mJ K$ ^{-1}$cm$^{-3}$ and hold time below 2~K of 220~min are competitive while the minimal temperature is higher compared to frustrated Ytterbium-oxide ADR materials studied under similar conditions.
\end{abstract}
\maketitle


\section{Introduction}
Ultra-low temperatures below 1~K are often prerequisites for the occurrence of quantum phenomena in condensed matter and, respectively, key for modern quantum technologies.
The global helium shortage~\cite{kelley8,olafsdottir1,kouzes2010,shea2011,Cho778} and the rising need for milli-Kelvin (mK) refrigeration lead to a resurrection of adiabatic demagnetization refrigeration (ADR), a long-established helium-free cooling technique~\cite{debye1154, giauque768,pobell2007}.
ADR cooling has the advantage of having less environmental impact and saving almost one-third of energy~\cite{Alahmer4662}, and it can be used in various applications such as quantum computation~\cite{jahromi2019}, space technologies~\cite{shirron130,shirron581,shirron915,luchier591,hagmann303}, etc. 
For mK-ADR applications, materials with a significant entropy change (in the temperature range of 20~mK to 5~K) and a strong magneto-caloric effect (MCE)~\cite{weiss103} (in a minimal applied magnetic field) are preferred. In ADR, the coolant material is pre-cooled in a magnetic field of a few teslas and then adiabatically demagnetized. Due to the low entropy of the material after pre-cooling in the field, adiabatic demagnetization requires a further temperature reduction to mK. Recent advances in continuous mK-ADR address the ineffectiveness of ADR compared to $ ^{3} $He-$ 4 $He dilution refrigeration for continuous refrigeration~\cite{shirron581,bartlett582}. Commercial continuous refrigerators based on ADR are already available on the market, suggesting their potential as a significant refrigeration advancement~\cite{kiutra}.

For cooling down to mK, hydrated paramagnetic salts such as CrK(SO$ _4$)$ _2$$\cdot$12H$ _2$O (CPA), which orders at $30$~mK, are typically used in commercial ADR setups~\cite{daniels243}. However, the necessary dilution of magnetic moments in hydrated paramagnetic salts drastically reduces the volumetric cooling power. Conversely, materials with higher concentrations of magnetic ions may exhibit enhanced exchange couplings, leading to magnetic order. The presence of magnetic ordering hinders further adiabatic refrigeration, as it reduces entropy towards zero and sets a temperature limit for ADR~\cite{tokiwa1}. While an ideal refrigerant would be a perfect paramagnet with maximum entropy at zero-field around 0~K, actual paramagnetic substances inevitably have weak interactions that affect their refrigeration performance by limiting the minimal ADR temperatures. In addition, hydrated organic paramagnetic salts need vacuum-tight encapsulation due to high water vapor pressure and there is a limitation in properly baking them above $ 100~^\circ$C for UHV uses. Additionally, they have very poor intrinsic thermal conductivity, which requires the elaborate incorporation of metal wires into the encapsulated ADR pills~\cite{wikus150}.

To overcome the disadvantages of hydrated paramagnetic salts for mK-ADR, different other material classes were explored. These include frustrated~\cite{zhitomirsky104421,hu125225,brasiliano103002,tokiwa1,liu1,Kleinhans014038,Jesche23,arjun2023}, quantum critical~\cite{wolf6862,wolf142112} and molecular magnets~\cite{evangelisti6606,baniodeh1}, as well as metallic rare-earth compounds~\cite{tokiwa1600835,jang1}. Among them, geometrically frustrated Ytterbium-oxides like KBaYb(BO$_3$)$_2$~\cite{tokiwa1} and (Na/K)YbP$_2$O$_7$~\cite{arjun2023}, combine several advantages: minimal ADR temperatures well below 50~mK, in combination with larger volumetric entropy density compared to hydrated paramagnetic salts, and, most importantly, chemical stability. ADR pellets based on these materials can easily be prepared by pressing polycrystals with fine silver powder (to ensure optimal thermal contact at low temperatures) and upscaled to the kilogram range.


As a result of spin orbit coupling and crystal electric field (CEF) splitting of the Yb$^{3+}$ states in an octahedral environment, a Kramers doublet with effective spin 1/2 moments arises in these materials at low temperatures. The effect of geometrical frustration in triangular rare-earth magnets has recently been investigated in YbMgGaO$_{4}$ and AYb$X_{2}$ ($A$=alkali metal, $X=$ Se, S, O)~\cite{Kimchi031028, Li097201, Paddison117,Bordelon1058,Ranjith224417,Baenitz220409,Schmidt214445,Ranjith180401}. Absence of long-range order together with broad excitation continua was related to the effect of geometrical frustration. In addition, the structural randomness of Mg$^{2+}$ and Ga$^{3+}$ in the former material gives rise to broad CEF splittings and broad distributions of the g-factors and exchange interactions. The structural randomness is therefore detrimental to long-range order and could mimic spin liquid behavior~\cite{Zhu2017}. KBaYb(BO$_3$)$_2$, with 40\% larger Yb-Yb distance compared to YbMgGaO$_{4}$ and no direct Yb-O-Yb links, features similar structural randomness of K$^+$ and Ba$^{2+}$ atoms, occupying the same position in the crystal lattice. This randomness could, together with frustration, further enhance the ADR performance~\cite{tokiwa1}. For isostructural KBaGd(BO$_3$)$_2$, which orders magnetically at 0.26~K, a minimal temperature of 0.12 K was obtained~\cite{Jesche23}. Indeed the phase transition in this material is strongly broadened and of triangular instead of $\lambda$-type shape, implying that the entropy release has been shifted by structural randomness towards much lower temperatures. More generally, the previous results motivate the investigation of other than triangular lattices without structural randomness with respect to their mK-ADR performance.


In the spin-1/2 2D frustrated square lattice (FSL) model or $J_1$-$J_2$ model, frustration may arise due to competing nearest-neighbor $J_1$ (along the edge) and next-nearest-neighbor $J_2$  (along the diagonal) interactions~\cite{Dagotto2148, Darradi214415, Jiang024424, Wang107202, Shannon027213, shannon599}. If both $J_1$ and $J_2$ are antiferromagnetic (AF), frustration arises, leading to interesting phases. The spin-1/2 FSL exhibits a rich theoretical phase diagram with various ground states depending on $J_1$ and $J_2$ or the frustration parameter $\alpha = J_1 / J_2$~\cite{shannon599, Mustonen064411}. The model exhibits classical ground states characterized by ferromagnetic (FM) order, N\'{e}el AF order, and columnar AF order. At the boundary between the two latter a QSL state has been predicted for $ J_2/J_1 \simeq 0.4$ to $0.6 $. Furthermore, a spin nematic phase is predicted between the FM and columnar AF phase~\cite{shannon599, Mustonen064411}. Although there are several spin-1/2 FSL candidate systems based on transition metals that fit well within the theoretical phase diagram~\cite{shannon599, Mustonen064411}, there are yet no good examples of rare-earth-based spin-1/2 FSL candidate systems.

In this paper, we report a thorough investigation down to mK temperatures on NaYbGeO$_{4}$, with effective 1/2 spins on a structurally ordered distorted square lattice. We do not find evidence for magnetic frustration. Instead the material displays a sharp $\lambda$-type phase transition in the heat capacity at 0.21~K. Indeed the minimal ADR temperature after demagnetization to zero field under similar conditions to previous studies on KBaGd(BO$_3$)$_2$~\cite{Jesche23} and (Na/K)YbP$_2$O$_7$~\cite{arjun2023} is only 0.15~K, while the warm-up time is enhanced compared to that of the two latter systems, as could also be expected from the $\sim 60\%$ higher entropy density. Altogether, the comparison supports the previous conjecture, that magnetic frustration and randomness are beneficial for optimized ADR performance.

\begin{figure*}
	\includegraphics[scale=0.5]{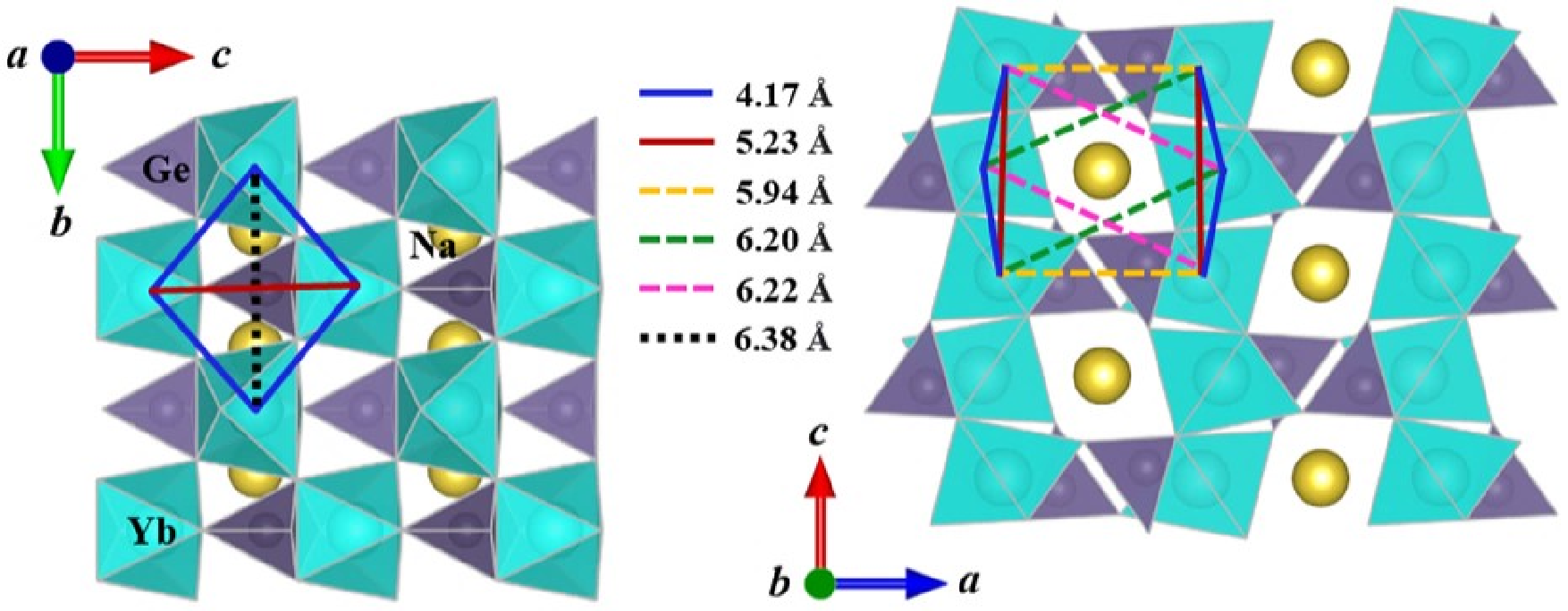}
	\caption{Crystal structure of NaYbGeO$_4$. The YbO$_{6}$ octahedra are linked via GeO$_{4}$ tetrahedra and span a distorted square lattice grid (left panel). Right panel shows the possible connections between the square lattice grids. Different Yb-Yb distances are color-coded.}
	\label{Fig1}
\end{figure*}

\section{Methods}

\textbf{Synthesis:} 
Polycrystalline samples of NaYbGeO$ _{4}$ were synthesized by solid-state reaction, annealing the stoichiometric mixture of Na$_2$CO$_3$ (99.99\%), Yb$_2$O$_3$ (99.99\%), and GeO$ _{2} $ (99.99\%) in an alumina boat at $ 1000 $~$^\circ$C for NaYbGeO$_{4}$ for a duration of $48$~h with one intermediate grinding and pelletization. 

\textbf{Powder X-Ray diffraction:} 
The phase purity of the samples was confirmed by powder x-ray diffraction (XRD, PANalytical powder diffractometer with Cu $K_{\alpha}$ radiation, $\lambda_{\rm ave} = 1.54182$~\AA) at room temperature. Rietveld refinement of the observed XRD patterns was performed using the \verb|FullProf|~package~\cite{Carvajal55} (see Figure~\ref{Fig2}), taking the initial parameters from Ref.~\onlinecite{Mehtap281}.

\textbf{DC magnetization:} 
DC magnetization ($M$) was measured as a function of temperature $T$ and applied magnetic field $H$ using the SQUID magnetometer with He-3 option [MPMS-3, Quantum Design].

\textbf{Specific heat:}
Specific heat ($C_{\rm p}(T)$), was measured using the heat capacity option of the PPMS (Quantum Design). For the low temperature ($ 0.4 $~K~$\leq T \leq 2.2$~K) $C_{\rm p}$ measurements, the $^{3}$He probe (Quantum Design) was utilized in the PPMS. Furthermore, we calculated the heat capacity below 0.4 K from the warming curve in the ADR experiment (see below).

To achieve strong thermal coupling, pellets made with NaYbGeO$ _{4}$ (grain diameter $10-50~\mu $m) and fine silver powders (grain diameter $1~\mu $m) in a mass ratio of $1:1$ were used for specific heat measurements. To obtain the specific heat of the sample, the contribution of silver was subtracted from the raw data. The specific heat $C_{\rm p}$ in a magnetic insulator has significant contributions from phonons ($C_{\rm ph}$) and the magnetic moments ($C_{\rm m}$). At high temperatures, $C_{\rm p}(T)$ is entirely dominated by $C_{\rm ph}$, while at low temperatures, it is dominated by $C_{\rm m}$. For estimating the $C_{\rm ph}$, the zero-field data were fitted by a polynomial function $C_{\rm ph}(T) = aT^3 + bT^5 + cT^7$ above $ 10 $~K, where $a$, $b$, and $c$ are appropriate constants~\cite{SM}. 
Similar procedures have been adopted earlier and proven to be efficient in cases where heat capacity data of non-magnetic analogue compounds are unavailable~\cite{Guchhait104409,Nath054409,Matsumoto9993,Lancaster094421}. The fit was extrapolated down to low temperatures, and $C_{\rm m}$ was obtained by subtracting the fitted data from the experimental $C_{\rm p}$ data. The magnetic entropy was estimated as $S_{\rm m}(T) = \int_{0.4~K}^{T}\frac{C_{\rm m}(T')}{T'}dT'$.

\textbf{Adiabatic Demagnetization Refrigeration (ADR):}
A home-made ADR setup for the PPMS was utilized~\cite{tokiwa1,Jesche23,arjun2023}. A cylindrical pellet made from equal weights of NaYbGeO$ _{4} $ (grain diameter $ 10–50~\mu $m) and silver powder (grain diameter $ 1~\mu $m) was pressed weighing approximately $3.5 $~g, with $ 15 $~mm diameter and $ 6 $~mm thickness. The pellet was mounted on a plastic straw and thermally isolated from the heat bath. A RuO$_{2}$ thermometer was attached to the pellet and connected through thin resistive manganin wires to minimize the heat flow. The resistor was measured with a current of $ 1 $~nA utilizing a Lake-Shore $ 372 $ AC Bridge. To minimize the effects of surrounding thermal radiation, a metallic cap was utilized as a shield. The sample was cooled to $T\approx 2$~K at $ 5 $~T and then the high-vacuum mode (pressure $< 10^{-4}$~mbar) was applied to achieve thermal decoupling. Subsequently, the magnetic field was swept to zero at a rate of $ 0.15 $~T min$ ^{-1} $. The pellet reached its lowest temperature and then slowly warmed back to $ 2 $~K via slow heat flow from the bath.

\section{Results}
\subsection{Structure}
The compound NaYbGeO$ _{4} $ crystallizes in the orthorhombic crystal structure with space group $ P n m a $ (No. 62). The crystal structure of NaYbGeO$ _{4} $ contains a distorted diagonally connected square lattice arrangement of YbO$ _{6} $ octahedra carrying the magnetic moments connected through GeO$ _{4} $ tetrahedra along the $ bc $-plane (see left panel of Fig.~\ref{Fig1}). The possible interaction pathways between these square lattice layers are shown in right panel of Fig.~\ref{Fig1}. From the structure, the nearest neighbor exchange is through corner-shared YbO$_6$ octahedra, while the second neighbor exchange is through GeO$_4$ tetrahedra. Both distances are smaller than $5.5$~Å, beyond which typically dipolar coupling dominates in other Ytterbium-oxides~\cite{Jesche23}.

\subsection{Powder X-Ray diffraction}
\begin{figure}
	\includegraphics[scale=0.22]{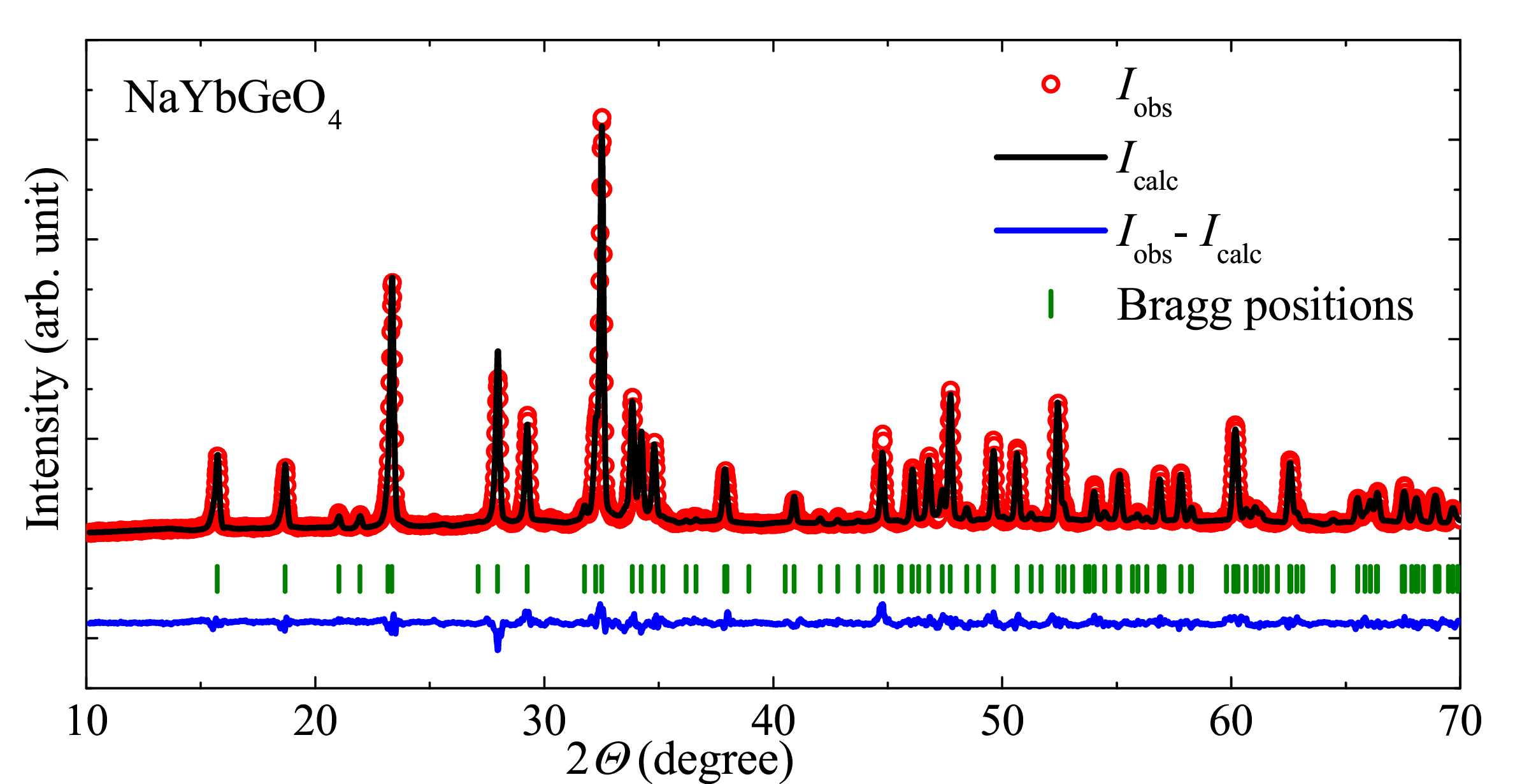}
	\caption{Powder x-ray diffraction pattern (open red circles) for NaYbGeO$_4$ at room temperatures. The solid line represents the Rietveld refinement, with the vertical bars showing the expected Bragg peak positions and the lower solid blue line representing the difference between observed and calculated intensities.}
	\label{Fig2}
\end{figure}

Figure~\ref{Fig2} shows the powder XRD pattern of NaYbGeO$ _{4} $ at room temperature, along with the Rietveld refinement. The refinement is done using the orthorhombic space group $Pnma$ (No. 62), taking initial parameters from Ref.~\onlinecite{Mehtap281}. The goodness of the fit is $ \chi^2 = 4.19  $. The obtained lattice parameters are $a = 11.2633(5)$~\AA, $b = 6.3756(3)$~\AA, $c = 5.2344(2)$~\AA, and $V_{\rm cell} \simeq 375.89$~\AA$^3$. These values are in close agreement with the reported values ~\cite{Mehtap281}.
\subsection{Magnetization}
\begin{figure}
	\includegraphics[scale=0.32]{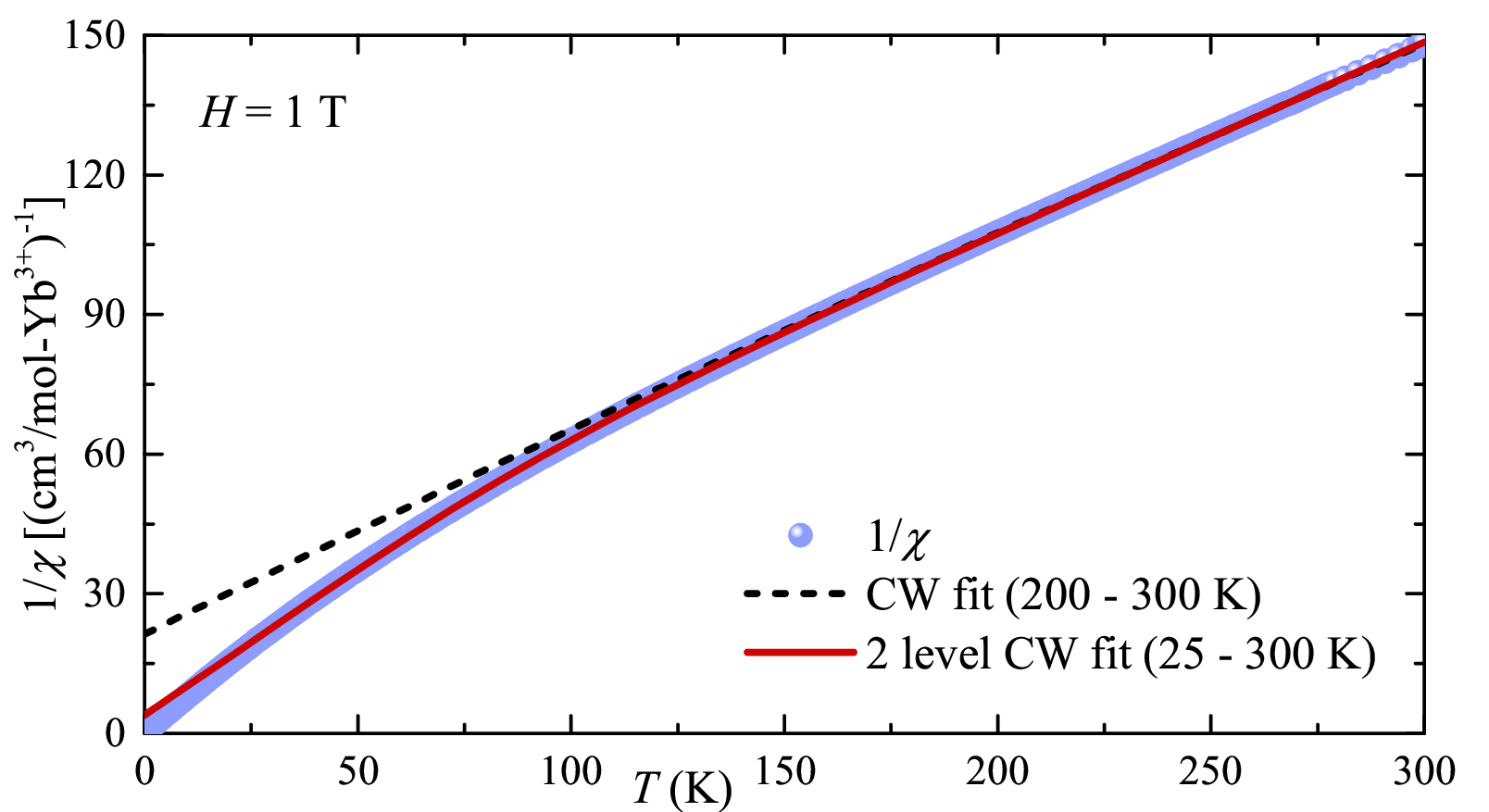}
	\caption{Inverse magnetic susceptibility $1/\chi$ vs. temperature of NaYbGeO$ _{4} $ at a field of $ 1 $~T. Dashed and solid lines represent the fits by Eq.~\ref{cw} and Eq.~\ref{2lev_cef}, respectively.}
	\label{Fig-MT}
\end{figure}

\begin{figure}
	\includegraphics[scale=0.32]{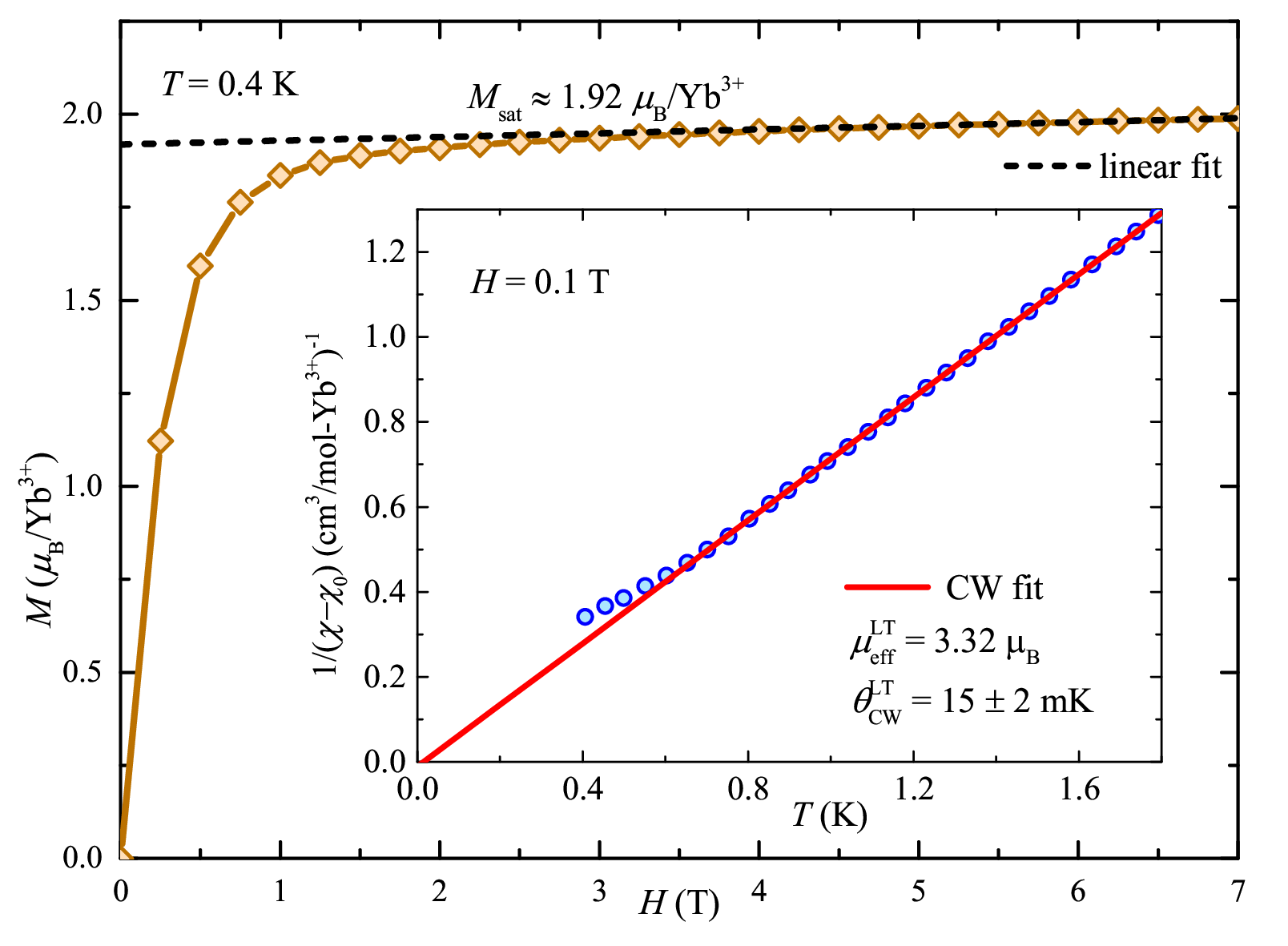}
	\caption{Isothermal magnetization $ M (H)$ of NaYbGeO$_4$  at a temperature of $ 0.4 $~K. Dashed line represents linear contribution (see text). The inset shows the low-temperature inverse magnetic susceptibility after subtracting $ \chi_{\rm 0} $ along with the CW fit.}
	\label{Fig4}
\end{figure}

Fig.~\ref{Fig-MT} displays the inverse magnetic susceptibility $\chi(T)^{-1}= H/M$ of NaYbGeO$ _{4} $ at 1 T field together with an extended Curie-Weiss (CW) fit according to 
\begin{equation}\label{cw}
	\chi(T) = \chi_0 + \frac{C}{T - \theta_{\rm CW}},
\end{equation}
which describes very well the high-$ T $ region between 200 and 300~K. Here, $\chi_0$ is the $T$-independent contribution consisting of the diamagnetic susceptibility ($\chi_{\rm core}$) of core electron shells and the Van-Vleck paramagnetic susceptibility ($\chi_{\rm VV}$) of the open shells of the Yb$^{3+}$ ions. The second term in Eq.~\ref{cw} is the CW law with the CW temperature $\theta_{\rm CW}$ and Curie constant $C = N_{\rm A} \mu_{\rm eff}^2/3k_{\rm B}$. Here, $N_{\rm A}$ is Avogadro's number, $\mu_{\rm eff} = g\sqrt{J(J+1)}$$\mu_{\rm B}$ is the effective magnetic moment, $g$ is the Land$\acute{\rm e}$ $g$-factor, $\mu_{\rm B}$ is the Bohr magneton, and $S$ is the spin quantum number. The fitting parameters are $\chi_{0}^{\rm HT} \simeq 4.34 \times 10^{-4}$ cm$ ^{3} $/mol, $ C_{\rm HT} \simeq 2.19$~cm$ ^{3} $K/mol, and $ \theta_{\rm CW}^{\rm HT} \simeq -47.1$~K. From the value of $ C_{\rm HT}$, the effective moment is obtained to $\mu_{\rm eff}^{\rm HT} \simeq 4.19 \mu_{\rm B}$, in good agreement with the expected spin-only value of $ 4.54 $~$\mu_{\rm B}$ for Yb$ ^{3+} $ ($ J = 7/2 $, $ g = 8/7 $) ion in the $ 4f^{13} $ configuration.The large value of the CW temperature arises from the CEF splitting. This is evident from the curvature in $\chi^{-1}(T) $ below $ 200 $~K leading to a change in slope below 50~K, as excited CEF levels become thermally depopulated. Such behavior is typical for Yb$^{3+}$-based spin systems where CEF splits the $J=7/2$ multiplet into four Kramers doublets.
Our entropy analysis below confirms the presence of a Kramers doublet ground state at temperatures below $10$~K.
Fig.~\ref{Fig-MT} also includes a two-level fit~\cite{mugiraneza95} (cf. the red solid line) according to 
\begin{equation}\label{2lev_cef}
	1/\chi(T) = 8 (T-\theta_{\rm CW}) \left(\frac {\mu_{\rm eff,1}^2 +\mu_{\rm eff,2}^2 \cdot e^{(-\Delta/T)}} {1+e^{(-\Delta/T)}}\right)^{-1},
\end{equation}
where $\mu_{\rm eff,1}$ and $\mu_{\rm eff,2}$ correspond to energy levels separated by an energy gap $\Delta $.
The fit down to $ 25 $~K yields $\Delta\simeq 282$~K, $\mu_{\rm eff,1}\simeq 3.57~\mu_{\rm B}$, $\mu_{\rm eff,2}\simeq 5.11~\mu_{\rm B}$, and $\theta_{\rm CW}\simeq -6.1$~K. The deviation of the fit from the data below $ 30 $~K indicates that the obtained $\mu_{\rm eff,1}$ and $\theta_{\rm CW}$ do not represent the lowest Kramers doublet. Using a fitting approach that incorporates four CEF levels and their associated effective moments is challenging due to the excessive number of adjustable parameters. Hence, we analyzed the magnetic susceptibility below $2$~K separately using Eq.~\ref{cw} to determine the properties associated with the Kramers doublet ground state.

The magnetization isotherm $M(H)$ of NaYbGeO$_{4}$ measured up to 7~T at $T=0.4$~K shows a saturation at around $ 1.5 $~T (Fig.~\ref{Fig4}). The saturation value $M_{\rm sat}$ is estimated by fitting the high-field region above $ 5 $~T with a straight line reflecting $\chi_0 H$ and extrapolating the line back to zero field. This yields $\chi_0 \simeq 5.58\times 10^{-3}$~cm$ ^{3} $/mol and $M_{\rm sat} \simeq 1.92~\mu_{\rm B}$, giving $g_{\rm eff}\simeq 3.84 $ from $\mu_{\rm sat}=g_{\rm eff}S_{\rm eff}~\mu_{\rm B}$.

After subtracting $\chi_0$, the $\chi(T)$ data below $2$~K are fitted by the CW law $\chi(T)=C/(T-\theta_{\rm CW})$. The fitting yields $ \theta^{\rm LT}_{\rm CW} \simeq (15\pm 2)$~mK and $\mu_{\rm eff}^{\rm LT} \simeq 3.32~\mu_{\rm B}$. The obtained effective moment is in good agreement with $\mu_{\rm eff} = g_{\rm eff}\sqrt{S_{\rm eff}(S_{\rm eff}+1)}~\mu_{\rm B}$ for a pseudo spin-$ 1/2 $ ground state with $g_{\rm eff}$ value of $ \sim 3.84 $, consistent with the value estimated from $ M_{\rm sat} $. The CW temperature depends on the sum of all exchange interactions and thus the balance of AF and FM contributions could result in a very small value of $\theta^{\rm LT}_{\rm CW}$.
Analysis of ADR curves, detailed later, evidences AF order at 210~mK, which proves that the extremely small value of $\theta_{\rm {CW}}$ must arise from the sum of AF and FM contributions. Note that a combination of nearest neighbor AF and next-nearest neighbor FM interactions on the square lattice excludes significant frustration.

\subsection{Specific heat}

\begin{figure}
	\includegraphics[scale=0.32]{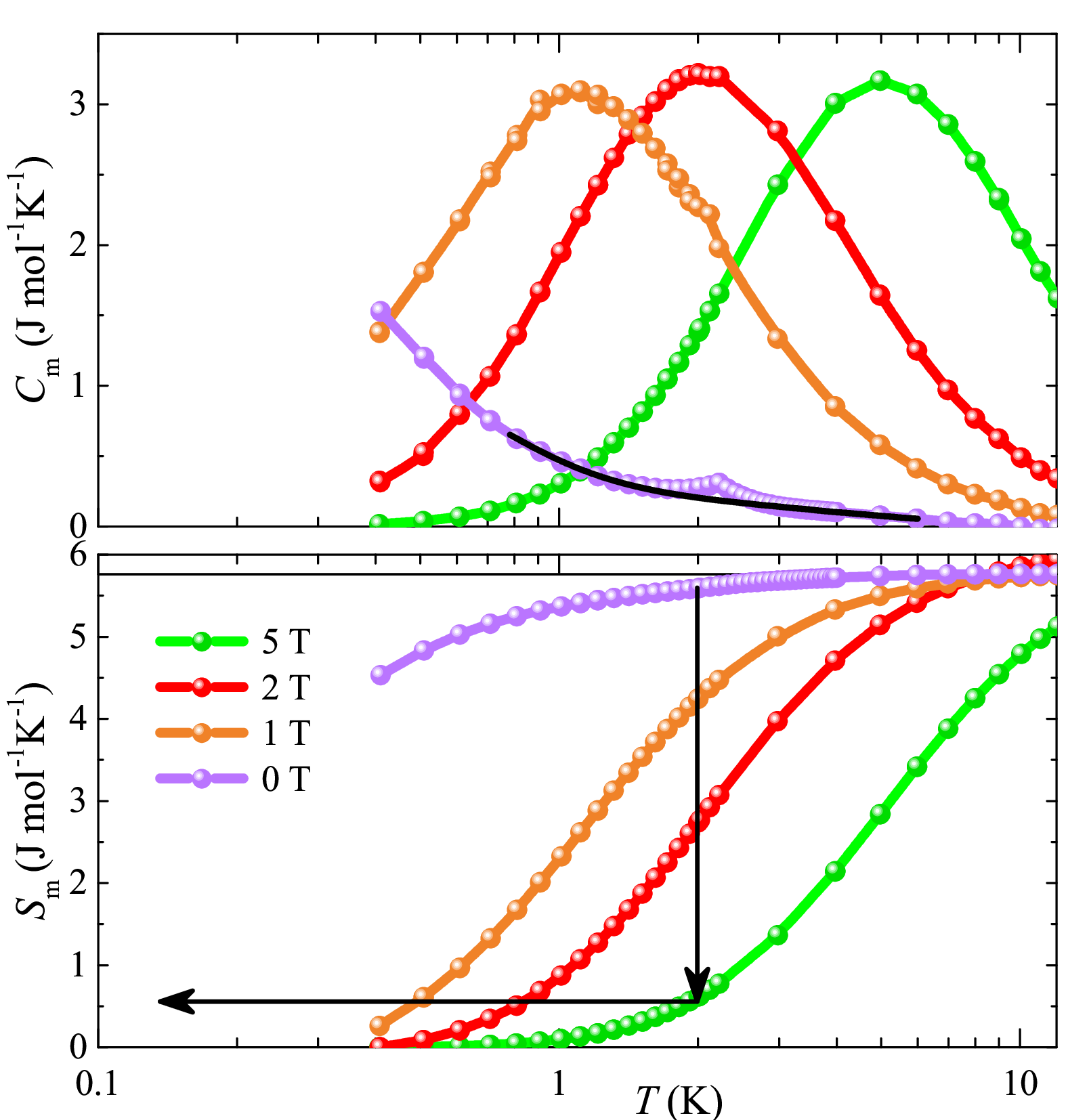}
	\caption{Low-temperature magnetic specific heat $ (C_{\rm m}) $ of NaYbGeO$_4$ at several external magnetic fields (upper panel). Phonon contribution was subtracted from the raw data using a polynomial fit~\cite{SM}. The lower panel displays the magnetic entropy $ S_{\rm m}(T)$ obtained by integrating $ C_{\rm m}/T $ over temperature. For the $ 0 $ and $ 1 $~T data the integration constant was adjusted such that a value of $R\ln 2$ (thin black line) is reached above 10~K. Two arrows show the ADR process.}
	\label{Fig7}
\end{figure}

Figure~\ref{Fig7}, upper panel, illustrates the magnetic specific heat for NaYbGeO$ _{4} $ under various applied fields, suggesting its potential for mK-ADR. The zero-field specific heat data show no indication of magnetic long-range ordering above $ 0.4 $~K. A small anomaly at 2.23~K, cf. the difference between the data and the solid blue line, arises from $0.6\%$ Yb$_2$O$_3$ impurities~\cite{SM}. The finite-field data show Schottky anomalies, resulting from the splitting of the lowest Kramers doublet in field. This is confirmed by the fact that the entropy approaches $R\ln 2$, as displayed in the lower panel. As indicated by the two arrows, starting the ADR at $\sim 2$~K in a field of 5~T should reveal a minimal temperature well below 0.4~K in zero field.



\subsection{Adiabatic Demagnetization Refrigeration (ADR)}
\begin{figure}
	\includegraphics[scale=0.32]{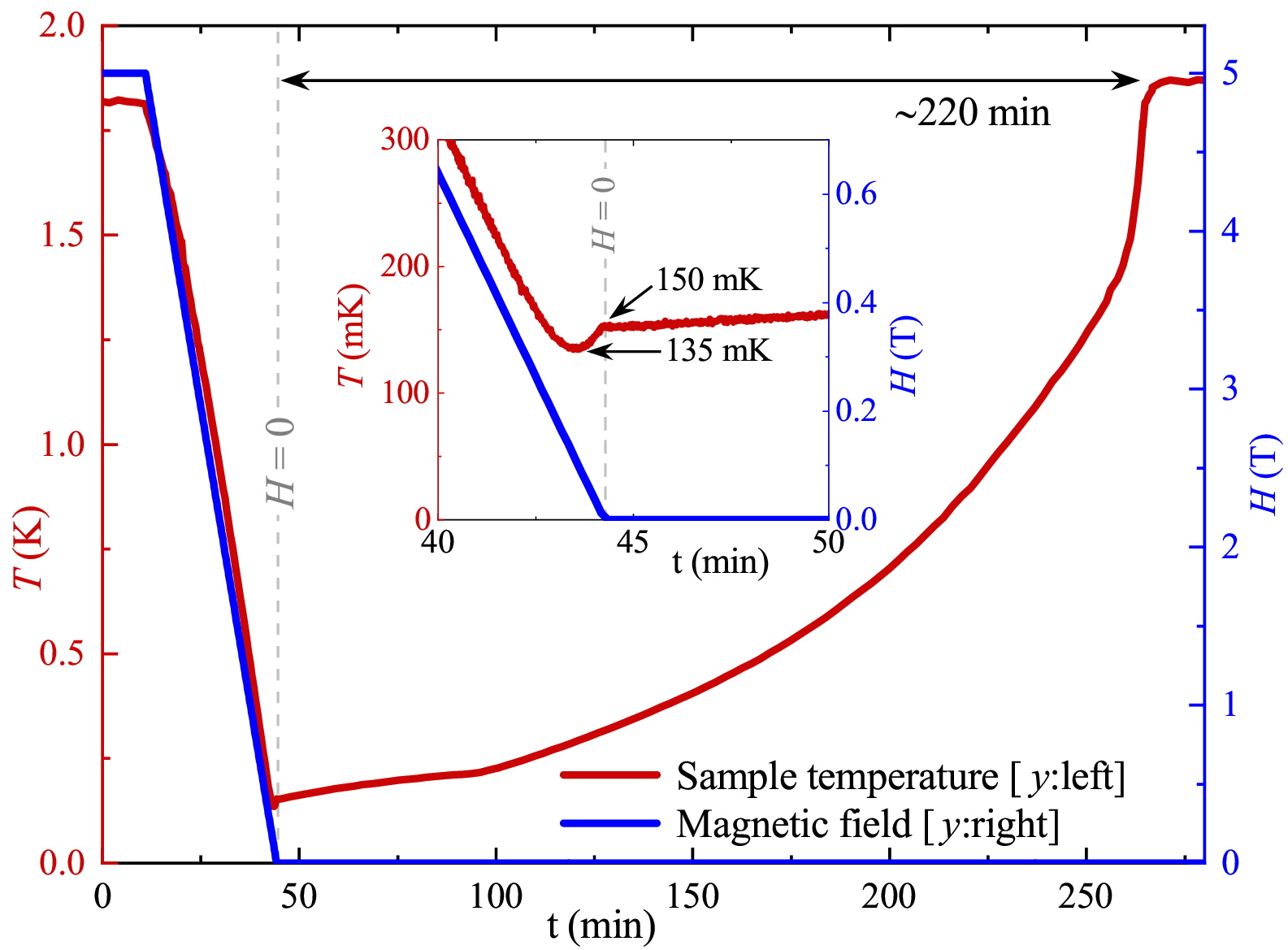}
	\caption{Cooling of NaYbGeO$ _{4} $ by ADR. The red and blue lines display the temporal evolution of temperature (left y-axis) and magnetic field (right y-axis). After precooling in a field of 5~T the sample space was evacuated and subsequently the magnetic field was swept from $ 5 $ to $ 0 $~T at a rate of $ 0.15 $~T min$ ^{-1} $. The inset enlarges the regime close to zero-field, where the temperature reaches a minimum of 135~mK at 0.1~T and then saturates at 150~mK in zero field. The warm-up time amounts to 220~min.}
	\label{Fig11}
\end{figure}
\begin{figure}
	\includegraphics[scale=0.32]{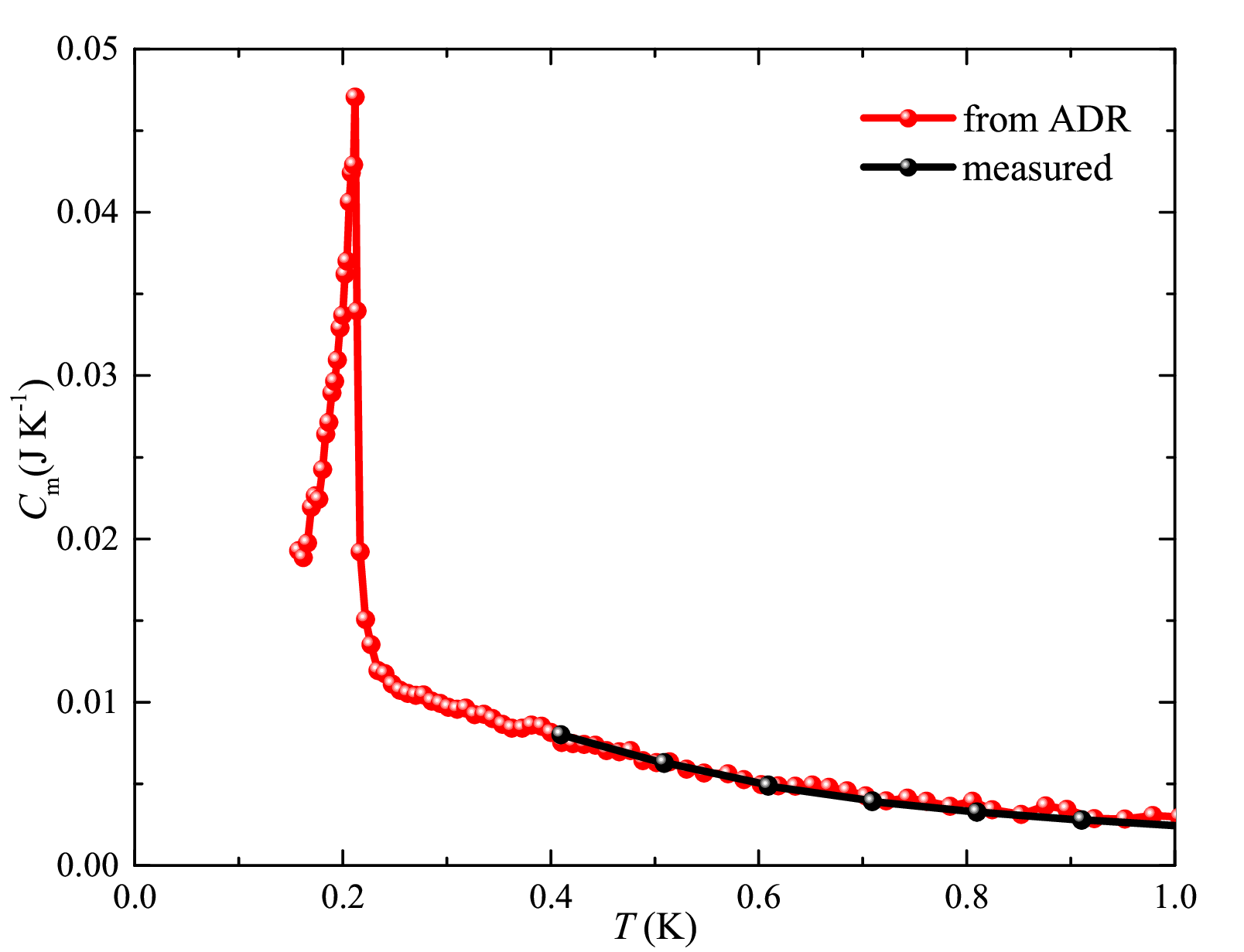}
	\caption{Temperature dependence of the total (magnetic) heat capacity of the NaYbGeO$_4$-based ADR pellet determined from the warming curve shown in Fig~\ref{Fig11} by calculating $C_{\rm ADR}=\dot Q/\dot T$ (red points). The constant heat input $\dot Q=0.58 \mu$W was used to obtain agreement with the measured heat capacity (black points) above 0.4~K.}
	\label{Fig_Cm_ADR}
\end{figure}

The cooling performance of NaYbGeO$ _{4} $ in our PPMS setup described above is shown in Fig~\ref{Fig11}. Near 0.1~T the temperature trace passes a minimum at 135~mK. This sign change of the adiabatic magnetocaloric effect indicates the crossing of a magnetic phase boundary. Upon further reducing the field, the temperature increases towards 150~mK in zero field. The subsequent warming results from the finite heat input due to non-perfect adiabatic conditions. The time dependence of the temperature $T(t)$ displays a clear anomaly near 100 minutes. To analyze this trace, we utilize 
the equation
\begin{equation}\label{Cm_ADR}
	\dot Q = C_{\rm ADR} \cdot \dot T,
\end{equation}
where $\dot{Q}$ represents a (constant) heat input per unit time, $ C_{\rm ADR} $ denotes the (magnetic) heat capacity of the ADR pellet and $\dot{T}$ is the time derivative of temperature during warming.
Taking $\dot Q = 0.58 \mu$W, a perfect agreement between the directly measured heat capacity and $C_{\rm ADR}$ is found down to $0.4$ K. This heat input mainly arises from the residual He gas in our setup and here it was slightly lower than the $0.71 \mu$W determined previously for KBaGd(BO$_{3}$)$_{2}$~\cite{Jesche23}. For KBaGd(BO$_{3}$)$_{2}$ we also confirmed that the calculated heat capacity exactly equals the measured heat capacity signature of the AF transition. This provides us strong confidence that the sharp transition in $C_{\rm ADR}$ for NaYbGeO$_4$ distinctly signifies sharp AF order at $T_{\rm N}\simeq 210$ mK. Note, that the sign change of the magnetocaloric effect in the field of 0.1~T at 130~mK is compatible with the suppression of the AF order in field.

The warm-up time of $\sim 220$~min (3~hr 40~min) is approximately 4-6 times longer than that of KYbP$ _{2} $O$ _{7} $ ($\sim35$~min), NaYbP$ _{2} $O$ _{7} $ ($\sim55$~min), KBaYb(BO$ _{3} $)$ _{2} $ ($\sim40$~min)~\cite{arjun2023} though much less than $\sim $8~hr for KBaGd(BO$ _{3} $)$ _{2} $~\cite{Jesche23}. This reflects that the entropy density 101~mJ/(K cm$^3$) of NaYbGeO$_4$ is roughly 60\% larger than that of the former materials but only half of that in KBaGd(BO$_{3}$)$_{2}$. The latter material reaches a similar (even slightly lower) ADR end temperature of 122~mK~\cite{Jesche23} compared to 135~mK in NaYbGeO$_4$ and is therefore even better suitable for applications in this target range. 

%

\section{Discussion}

\begin{table}[h]
	\caption{Comparison of important parameters of different mK ADR materials: $T_{\rm N}$ is the magnetic ordering temperature, $T_{\rm min}$ is the minimum temperature attained, $S_{\rm GS}$ is the entropy of the ground-state multiplet, and $R$ is the universal gas constant. The abbreviations are MAS = Mn(NH$_{4}$)$_{2}$(SO$_{4}$)$_{2}\cdot6$H$_{2}$O (manganese ammonium sulfate), FAA = NH$_{4}$Fe(SO$_{4}$)$ \cdot $$12$H$_{2}$O (ferric ammonium alum), CPA = KCr(SO$_4$)$ \cdot 12$H$_{2}$O (chromium potassium alum), CMN = Mg$_{3}$Ce$_{2}$(NO$_{3}$)$_{12}\cdot24$H$_{2}$O (cerium magnesium nitrate).}
	\label{ADR_parameters}
	\begin{tabular}{ccccc}
		\hline \hline
		&&ADR materials&&\\
		\hline
		Material & $ T_{\rm N} $ & $ T_{\rm min} $& $ S_{ \rm GS}$ & $ S_{\rm GS} $/vol. \\
		& (mK) & (mK) &  & [mJ/(K cm$ ^{3} $)] \\
		\hline
		MAS~\cite{Vilches509} & $ 170 $ & $ <300 $ & $R\ln 6$ & $ 70 $ \\
		FAA~\cite{Vilches509} & $ 30 $ & $ <100 $ & $R\ln 6$ & $ 53 $\\
		CPA~\cite{daniels243} & $ 10 $ & $ <100 $ & $R\ln 4$ & $ 42 $\\
		CMN~\cite{Fisher5584} & $ 2 $ & $ <40 $ & $R\ln 2$ & $ 16 $\\
		KBaGd(BO$_{3}$)$_{2}$~\cite{Jesche23} & $ 263 $& $ 122 $& $R\ln 8$ & $ 192 $\\
		YbPt$ _{2} $Sn$ _{12} $~\cite{jang1} & $ 250 $ & $ <200 $ & $R\ln 2$ & $ 124 $\\
		Yb$ _{3} $Ga$ _{5} $O$ _{12} $~\cite{brasiliano103002} & $ 54 $ & $ <200 $ & $R\ln 2$ & $ 124 $\\
		\textbf{NaYbGeO$_{4}$} & $\textbf{210}$ & $ \textbf{135} $ & $\textbf{\textit{R} ln 2}$ & $ \textbf{101} $\\
		KBaYb(BO$_{3}$)$_{2}$~\cite{tokiwa1} & $ 8 $& $ <20 $& $R\ln 2$ & $ 64 $\\
		NaYbP$ _{2}$O$_{7}$~\cite{arjun2023} & $ <45 $ & $ <45 $ & $R\ln 2$ & $ 64 $\\
		KYbP$ _{2}$O$_{7}$~\cite{arjun2023} & $ <37 $ & $ <37 $ & $R\ln 2$ & $ 57 $\\
		\hline \hline
	\end{tabular}
\end{table}

The sharp AF phase transition at 210~mK indicates that NaYbGeO$_4$ with distorted square-lattice arrangement of Yb$^{3+}$ moments is much weaker magnetically frustrated compared to the triangular-lattices KBaYb(BO$_{3}$)$_{2}$~\cite{tokiwa1} and NaYbP$ _{2}$O$_{7}$~\cite{arjun2023}, as well as the distorted hyperhoneycomb-lattice KYbP$ _{2}$O$_{7}$~\cite{arjun2023}, which all show no magnetic order above 35~mK ($T_{\rm N}=8$~mK in KBaYb(BO$_{3}$)$_{2}$ has been deduced by comparison with KBaGd(BO$_{3}$)$_{2}$~\cite{Jesche23}).
The fact that the CW temperature in NaYbGeO$_4$ is much smaller compared to the ordering temperature could be attributed to AF nearest neighbor and FM second neighbor exchange couplings. In this configuration, no frustration arises, as both first and second neighbor interactions are consistently satisfied simultaneously. If any frustration exists, it would likely arise from interactions perpendicular to the plane. 

A comparison of the ADR performance of NaYbGeO$_4$ with that of other ADR materials, in particular geometrically frustrated Ytterbium-borates and -diphosphates is therefore interesting, as this may allow to draw conclusions on the importance of magnetic frustration in this respect.
The crucial parameters that determine the efficiency of the various known ADR materials are compared in Table~\ref{ADR_parameters}. The full entropy of the ground state is described as $ S_{\rm GS}= R\ln (2J+1) $, and the entropy density $ S_{\rm GS} $/vol. is calculated by dividing $ S_{\rm GS}$ by the unit cell volume. A large value of $ S_{\rm GS}$ is always beneficial because the magnetic entropy changes ($ \Delta S_{\rm m} $) of magnetic refrigerants act as the driving force of ADR. However, it is worth noting that for practical purposes it is not the molar entropy but the volumetric entropy density $S_{\rm GS} $ of the material that is relevant.

As can be seen from the Table.~\ref{ADR_parameters}, materials with high entropy density such as YbPt$_{2}$Sn or Yb$_{3}$Ga$_{5}$O$_{12}$ exhibit magnetic orderings at $ 250 $ mK and $ 54 $~mK, respectively, which limit their lowest attainable temperatures as the entropy drops below the ordering temperatures. The high spin hydrated paramagnetic salts such as MAS and FAA, which have a larger magnetic entropy $R\ln 6$ are also affected by their much higher magnetic ordering temperatures. On the other hand, low transition temperature materials such as CPA and CMN have low magnetic moment density and hence low entropy density. A high entropy density usually contradicts a low magnetic ordering temperature.

In this context, it has been reported that the frustrated Yb-based quantum magnets $ A $YbP$ _{2} $O$ _{7} $, KBaYb(BO$ _{3} $)$ _{2} $, exhibit these two mutually exclusive criteria, such as a high volumetric entropy density combined with a very low ordering~\cite{arjun2023, tokiwa1}, allowing ADR to well below 50~mK. In KBaYb(BO$ _{3} $)$ _{2} $ both magnetic frustration and structural randomness contribute to suppress magnetic ordering. These oxides have been shown to be excellent anhydrous ADR refrigerants and under the same experimental conditions in PPMS, out of these three compounds, KYbP$ _{2} $O$ _{7} $ is attaining the lowest temperature of $37$~mK~\cite{arjun2023}. Utilizing a better adiabatic setup in the $ ^{3} $He-$ ^{4} $He dilution refrigerator with feedback control of the bath temperature following the sample temperature, these materials can reach even lower temperatures well below $ 20 $~mK, which has been experimentally demonstrated for KBaYb(BO$ _{3} $)$ _{2} $~\cite{tokiwa1}.

NaYbGeO$ _{4} $ has a large magnetic ion density ($ \sim 10.6 $~nm$ ^{-3} $) and volumetric entropy density ($ \sim 101 $~mJ/(K cm$ ^{3} $)). These values are higher than that of $ A $YbP$ _{2} $O$ _{7} $ and KBaYb(BO$ _{3} $)$ _{2} $ and this is reflected in the significantly longer warm up time under comparable conditions. However, the end temperature of 135 mK is much higher compared to the case of the frustrated materials, indicating the importance of frustration for obtaining low end temperatures.


Compared to conventional mK ADR coolants based on hydrated paramagnetic salts, NaYbGeO$ _{4}$ is stable at high-vacuum and can be heated up to $ 1000~^\circ $C making it easy to handle and suitable for UHV applications.
For our studies we prepared ADR pellets by mixing powder samples with silver powder in $ 1:1 $ ratio for getting excellent thermal contact. The NaYbGeO$ _{4}$ pellets, sintered at $1000~^\circ $C, are mechanically stronger compared to the diphosphate pellets which we sintered only at $600~^\circ $C~\cite{arjun2023}. Thus, mechanically stable NaYbGeO$ _{4}$ ADR pills could also be made with much less or even without silver powder.

\section{Conclusion}
We synthesized and studied NaYbGeO$_4$ with effective spin 1/2 moments on a distorted square lattice with respect to its performance in milli-Kelvin (mK) adiabatic demagnetization refrigeration (ADR). While the Curie-Weiss temperature of 15~mK would suggest the absence of significant magnetic interactions, the heat capacity, derived from the warming curve after ADR shows a sharp AF phase transition at $T_{\rm N}\simeq 210$ mK. This suggests AF nearest neighbor and FM second neighbor exchange couplings and a non-frustrated magnetic state. Compared to geometrically frustrated Ytterbium-borates and -diphosphates the minimal ADR temperature of NaYbGeO$_4$ is significantly larger. This supports the conjecture that magnetic frustration is beneficial for mK-ADR.

Even though NaYbGeO$_{4}$ has a limitation in the lowest attainable temperature ($\sim 135 $~mK) due to the presence of sharp AF ordering, it has a comparably high entropy density and respectively shows a much longer warm-up time than Ytterbium-borates and -diphosphates. Chemical stability upon heating up to $ 1000~^\circ $C, UHV compatibility, and the easy production of mechanically stable sintered pellets make this material an interesting alternative to hydrated paramagnetic salts for ADR applications above 200~mK.

\acknowledgments
We thank Marvin Klinger and Yoshi Tokiwa for useful discussions. UA would like to acknowledge DST, India, for financial support bearing sanction (DST/INSPIRE/04/2019/001664). Work supported by the German Research Foundation through project 514162746 (GE 1640/11-1). DDS thanks SERB, DST, and CSIR, Government of India, for financial support.

\bibliography{ref_NYG}

\end{document}